\documentclass[aps,pre,twocolumn,groupedaddress,showpacs]{revtex4-1}

\usepackage[utf8]{inputenc}
\usepackage[english]{babel}
\usepackage{amsmath,graphicx,enumerate}
\begin{document}


\title{A kinetic model for the finite-time thermodynamics of small heat engines} 


\author{Luca Cerino}
\author{Andrea Puglisi}
\author{Angelo Vulpiani}
\affiliation{Istituto dei Sistemi Complessi - CNR and Dipartimento di Fisica, Universit\`a di Roma Sapienza, P.le Aldo Moro 2, 00185, Rome, Italy}


\date{\today}

\begin{abstract}
We study a molecular engine constituted by a gas of $N \sim
10^2$ molecules enclosed between a massive piston and a
thermostat. The force acting on the piston and the temperature of the
thermostat are cyclically changed with a finite period $\tau$. In the
adiabatic limit $\tau \to \infty$, even for finite size $N$, the
average work and heats reproduce the thermodynamic values, recovering
the Carnot result for the efficiency. The system exhibits a stall
time $\tau^*$ where net work is zero: for $\tau<\tau^*$ it
consumes work instead of producing it, acting as a refrigerator or as
a heat sink. At $\tau>\tau^*$ the efficiency at maximum power is close to the
Curzorn-Ahlborn limit. The fluctuations of work and heat display
approximatively a Gaussian behavior. Based upon kinetic theory, we
develop a three-variables Langevin model where the piston's position
and velocity are linearly coupled together with the internal energy of
the gas. The model reproduces many of the system's features, such as
the inversion of the work's sign, the efficiency at maximum power and
the approximate shape of fluctuations. A further simplification in the
model allows to compute analytically the average work, explaining its
non-trivial dependence on $\tau$.
\end{abstract}



\maketitle
\section{Introduction}

The usual statistical mechanics treats macroscopic objects containing
an enormous number $N$ of particles, at least $O(10^{20})$; the
classical thermodynamics refers to adiabatic processes. In practice a
transformation can be considered adiabatic if its typical times are
much longer than the times involved in the dynamics of the underlying
system.  Basically in the standard statistical mechanics and
thermodynamics two asymptotic limits are present: large $N$ and very
slow changes of parameters~\cite{Castiglione2008,Reichl2009}. The
challenge we face nowadays is going beyond these limits, extending
thermodynamics and statistical mechanics to new models and
applications~\cite{Gaspard2006}.

In fact, on one hand, it is clear that real transformations occur in
finite time: this problem has been frequently discussed in the recent
past, giving birth to the so-called finite time
thermodynamics~\cite{Andresen1977,Vandenbroeck2005}, which focuses on
the study of engines working at finite power, i.e. far from Carnot
efficiency. On the other hand the recent technological progresses
allows us to relax also the large $N$ limit: now it is possible to
manipulate even small systems (say few hundreds particles) with non
adiabatic changes of the parameters~\cite{Bustamante2005}.  Therefore
it is necessary to (re)consider in details some aspects which are not
particularly relevant for macroscopic bodies.  As an important example we
mention the progresses 
 in the study of
fluctuations and their relation with response
functions~\cite{marconi2008fluctuation}.

For the ambitious project of establishing a suitable statistical
mechanics (as well as thermodynamics) formalism for small systems and
non adiabatic processes, it is necessary to build a theoretical
framework, with new paradigmatic models, able to give an efficient
description of the statistical features at the mesoscale. The
prototype of such a description is the Langevin equation, which is
able to catch the behavior of a colloidal particle (an object between
the microscopic realm and the macroscopic one).  The original Langevin
equation has been established with a clever combination of macroscopic
arguments (the Stokes law for the friction force) and the use of
statistical properties (equipartition).  Following the Smoluchowski
approach to the Brownian motion, sometimes, it is possible to
rationalize the building of a Langevin equation.  For instance for a
big intruder in a diluted gas, using the kinetic theory, one can
derive the precise shape of the friction force~\cite{vanKampen2007}. 
A step forward in this direction is represented by stochastic
thermodynamics, based on the idea that the thermodynamical concepts of
work, heat and entropy can consistently be extended to a single
trajectory~\cite{Sekimoto2010,Seifert2011}.

The aim of the present work is the study of the thermodynamic 
properties (work, heat, efficiency) in small systems performing non
adiabatic cycles.  We focus on a model system composed of a gas of
molecules, a thermostat and a moving piston 
\cite{sekimoto02,sano14,sasa15}, proposing a particular transformation
(Ericsson cycle).  The model's results are compared with those
obtained for a coarse-grained stochastic equation which
describes the evolution of slowly changing quantities, such as
position and velocity of the piston and temperature of the gas. Such
a Langevin equation is obtained by means of kinetic theory
considerations, in the spirit of the Smoluchowski approach. 

The article is organized as follows: in Sec. \ref{sec: 1} the model 
and the engine protocol are presented; in Sec. \ref{sec: 2} 
the results of numerical MD simulations of the model are reported, 
focusing on the average work and heat, the efficiency and the
fluctuations of such quantities. In Sec. \ref{sec: 3} the 
coarse-grained stochastic model is derived and compared with the 
original system: finally, in Sec. \ref{sec: 4} a simpler model is 
derived from the previous one and, in such a context, some analytical 
results are obtained. The details of the calculations are discussed in 
the Appendices. 

\section{The Model} 
\label{sec: 1}
\begin{figure}[!htbp]
\centering
\includegraphics[width=0.4\textwidth]{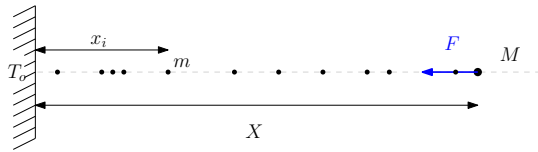}
\caption{Sketch of the piston model. A gas of particles is confined by 
a fixed wall (the thermal bath) and a moving wall (the piston) that is 
subject to a constant external force.}\label{fig: model}
\end{figure}

We consider an ideal gas of $N$ pointlike particles with mass $m$, 
position ${\bf x}_i$ and momentum ${\bf p}_i$ (${\bf v}_i={\bf 
p}_i/m$), $i=1\ldots N$, enclosed in a box. One of the sides of the 
box is a piston of mass $M$ and momentum $P$ which moves in the 
$\hat{x}$ direction. Indipendently of the real dimensionality of the 
box, only the motion in the $\hat{x}$ direction is relevant, as the 
particles interact only with the piston (see Fig. \ref{fig: model}).  
An externally controlled force $\vec{F}(t)=-F\hat{x}$ acts on the 
piston. The full (one-dimensional) hamiltonian of the system reads
\begin{equation}\label{eq: hamiltonian}
\mathcal{H}=\sum_{i=1}^N \frac{p_i^2}{2m}+\frac{P^2}{2M} +FX,
\end{equation}
with the additional constraints $0<x_i<X$, $i=1,\ldots,N$, and $X>0$. 
The collisions with the piston are assumed to be elastic
\begin{eqnarray}\label{eq: elastic}
V'&=&V+\frac{2m}{m+M}(v_i-V), \nonumber \\ 
v_i'&=&v_i+\frac{2M}{m+M}(V-v_i),
\end{eqnarray}
where $v_i'$ and $V'$ are post-collisional velocities. The wall at 
$x=0$ acts as a thermostat at the temperature $T_o$: a collision of a 
particle with the wall is equivalent to give a new velocity $v'$ to 
the particle with probability density
\begin{equation}\label{eq: thermostat}
\rho(v')=\frac{m}{T_o}v'e^{-\frac{m v'^{2}}{2T_o}}\Theta(v),
\end{equation}
where $\Theta(v)$ is the Heaviside Theta. Let us note that the 
presence of the piston introduces an interaction among the gas 
particles: for this reason the dynamics of the system depends on the 
number of particles $N$. As reported in previous studies of systems 
including the interaction between a piston and one or more gases 
\cite{lieb99,lesne06,cencini07,sasa15}, a relevant parameter for the 
dynamics is $Nm/M$: interesting behaviors are typically observed for 
values of this parameters $\mathcal{O}(1)$, as in our study. Hereafter 
we use arbitrary units in numerical simulations, and we put $k_B=1$ 
for the Boltzmann factor. At fixed $F$ and $T_o$, the study of the 
system in the canonical ensemble by means of standard statistical 
mechanics \cite{cerino2014fluctuations,Tfluct} reveals that $\langle X 
\rangle^{eq} =(N+1)T_o/F$ and$\sigma^2_X=(N+1)T_o^2/F^2$. In addition, 
if we define the estimate of the instantaneous temperature of the gas, 
\begin{equation}
T=\frac{1}{N}\sum_{i=1}^N m v_i^2,
\end{equation} 
the ensemble average of this quantity reads $\langle T\rangle^{eq}
=T_o$ and its variance is $\sigma^2_T=2\,T_o^2/N$.

\subsection{The engine protocol}
When the parameters $F$ and $T_o$ vary in time, mechanical work can be 
extracted from the system. In particular if we identify the 
(thermodynamical) internal energy of the system with the value of the 
hamiltonian, $E(t)=\mathcal{H}({\bf x}(t),t)$, and, in addition, we 
define the input power as
\begin{equation}
\dot{W}=\left.\frac{\partial \mathcal{H}}{\partial t}\right|_{{\bf x}(t)},
\end{equation}
conservation of energy simply reads $\dot{E}(t)=\dot{Q}(t)+\dot{W}
(t)$, where $Q$ is the energy absorbed from the thermal wall. For an 
hamiltonian $\mathcal{H}$ as in Eq. (\ref{eq: hamiltonian}) one gets 
$\dot{W}=X\dot{F}$. Let us remark that this formula is different from 
the one obtained in standard thermodynamics, $\dot{W}=F\dot{X}$: this 
is due to the fact that we included the energy of the piston in the 
internal energy of the system \cite{Jarzynski2007}.

\begin{figure}[!hbtp]
\includegraphics[width=0.4\textwidth]{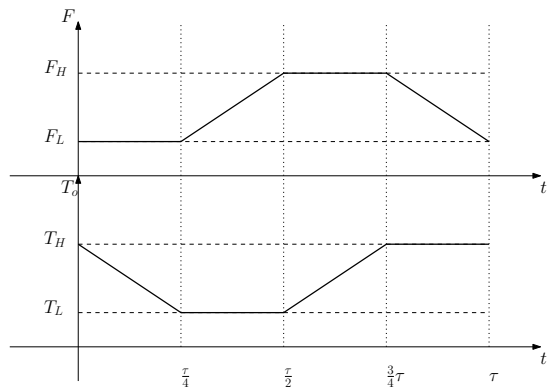}
\caption{Graph of $F$ and $T_0$ as a function of time over a cycle 
period $\tau$.}\label{fig: parameters}
\end{figure}

\begin{figure}
\centering \includegraphics[width=\columnwidth]{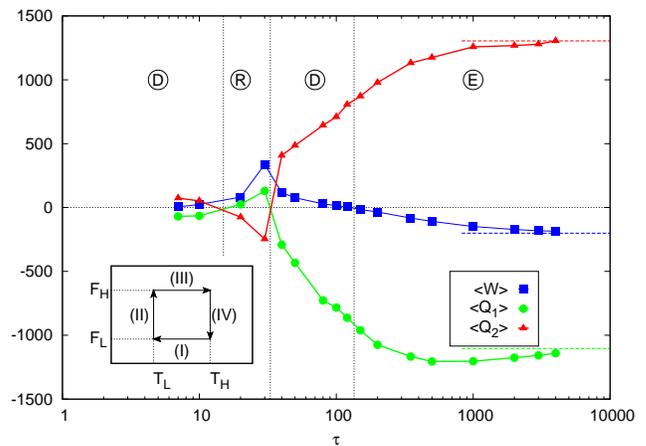}
\caption{Average values per cycle of work $W$ and heats $Q_1$, $Q_2$
as a function of the cycle typical time $\tau$. Dashed horizontal lines represent the adiabatic value of such quantities. Inset: schematic of the cycle protocol in the space of parameters $F,T_o$.  }\label{fig:ave}
\end{figure}

Here we adopt the following cyclical protocol to obtain a heat engine: 
the parameters vary periodically in time over a cycle of length $\tau$ 
(see Fig. \ref{fig: parameters} and the inset of Fig. \ref{fig:ave} 
for a visual explanation). If we set $t_0=k\tau$, with $k$ integer, 
the cycle has the following form:
\begin{enumerate}[I)]
\item At times $t \in
[t_0,t_0+\tau/4]$: isobaric compression ($F(t)=F_L$ and
$T_o(t)=T_H-4(T_H-T_L)(t-t_0)/\tau$),
\item  At times $t \in [t_0+\tau/4,
  t_0+\tau/2]$: isothermal compression
($F(t)=F_L+4(F_H-F_L)[t-(t_0+\tau/4)]/\tau$ and $T_o(t)=T_L$);
\item At times $t \in [t_0+\tau/2,t_0+3\tau/4]$ : isobaric expansion 
($F(t)=F_H$ and $T_o(t)=T_L+4(T_H-T_L)[t-(t_0+\tau/2)]/\tau$); 
\item  At times $t \in [t_0+3\tau/4,t_0+\tau]$: isothermal expansion
($F(t)=F_H-4(F_H-F_L)[t-(t_0+3\tau/4)]/\tau$ and $T_o(t)=T_H$).
\end{enumerate} The cycle
of length $\tau$ is repeated a large number of times over a long 
trajectory. In view of a Langevin-like analysis (see below) this 
protocol (called second type Ericsson cycle) -- which is thermostatted 
for the whole duration of the cycle -- is simpler than the more 
classical Carnot cycle.  A similar model has been studied 
in~\cite{Izumida2008} with the difference that the velocity of the 
piston is given and cannot fluctuate (infinite mass $M$ limit). The 
average values of heat and work  in each segment of the cycle can be 
 determined in the adiabatic limit, by substituting at every time 
$t$ the value of each variable with the equilibrium average: 
$X(t)=\langle X \rangle^{eq}_{F(t),T_o(t)}$ and $E(t)=\langle 
\mathcal{H}\rangle^{eq}_{F(t),T_o(t)}$ (see Table \ref{tab: 1}). 
\begin{table}[!htbp]
\centering
\begin{tabular}{c|c|c}
Segment&$\langle W\rangle$&$\langle Q\rangle$\\
\hline
I)&0&$\frac{3}{2}(N+1)(T_L-T_H)$\\
II)&$(N+1)T_L\ln\left(\frac{F_H}{F_L}\right)$&$-\langle W\rangle$\\
III)&0&$\frac{3}{2}(N+1)(T_H-T_L)$\\
IV)&$-(N+1)T_H\ln\left(\frac{F_H}{F_L}\right)$&$-\langle W\rangle$\\
\end{tabular}
\caption{Table with the adiabatic values of $Q$ and $W$ in each 
segment of the Ericsson cycle. The average $\langle \cdot\rangle$ is 
intended over many realization of the cycle.}\label{tab: 1}
\end{table}
In particular, let us note that on segments II) and IV) no work is 
performed and that the heats exchanged have same magnitude but 
opposite sign. Therefore, in the adiabatic limit, the system does not 
exchange net heat with any of the intermediate reservoirs at 
temperature $T_L<T^*<T_H$. Therefore in the rest of the paper we 
assume that two isobaric transformations do not contribute to the net
exchange of heat and work: this is true for $\tau \to \infty$ and
seems reasonable, for reasons of symmetry, at large $\tau$, while
(small) discrepancies at finite $\tau$ are observed in the
simulations. We will denote with $Q_1$ the heat exchanged with the 
cold reservoir $T_L$ in sector II), and with $Q_2$ the heat exchanged 
with  the thermostat at temperature $T_H$ in sector IV). If $\langle 
Q_2\rangle>0$ and $\langle W \rangle<0$, efficiency can be defined as
\begin{equation}
\eta=-\frac{\langle W \rangle}{\langle Q_2\rangle},
\end{equation}
where $W$ is the total work on a cycle, and $\langle \cdot \rangle$ 
denotes the average over many realizations of the cycle. Let us remark 
that this quantity is different from the average over many cycle of 
the fluctuating efficiency $\hat{\eta}= W/Q_2$.

\section{Results of MD simulations}
\label{sec: 2}
In order to perform molecular dynamics simulation of the system with 
time-dependent parameters $F$ and $T_o$, we introduce an interaction 
potential  between the piston and the particles to reproduce the 
effect of elastic collisions. We choose a repulsive soft sphere 
potential with cut-off radius $r_0$:
\begin{equation}
V(r)=\left[\left(\frac{r_0}{r}\right)^{12}+12\frac{r}{r_0}-13\right]\Theta(r_0-r),
\end{equation} 
where $\Theta(r)$ is the Heaviside Theta. The parameter $r_0$ is to 
be chosen as small as possible, compatibly with integration time-step 
$\Delta t$, in order to simulate a contact interaction. In our case 
$r_0=0.2$ and $\Delta t=0.0005$. The values of the other parameters, 
if not explicitly mentioned, are $m=1$, $M=100$, $N=500$, $T_L=11$, 
$T_H=13$, $F_L=180$, $F_H=220$. The integration scheme adopted is 
based on the standard Verlet algorithm.

\subsection{Average work and heats} 
 In Fig.~\ref{fig:ave} we report the
average values (over $500$ cycles) of $W$, $Q_1$ and $Q_2$ as a
function of $\tau$. In the adiabatic limit, $\tau \gg 10^3$, we 
recover, for $\langle W\rangle$, $\langle Q_1 \rangle$ and $\langle 
Q_2 \rangle$, the values computed assuming quasi-static 
transformations in thermodynamics. At finite values of the cycle's
duration $\tau < 10^3$ quite a complex scenario emerges. The absolute 
value of $\langle W \rangle$ decreases upon reducing $\tau$, until it 
vanishes at a stall time $\tau^* \sim 150$. For shorter cycles, the 
engine consumes work instead of producing it (the regime at 
$\tau>\tau^*$ is marked, on Fig.~\ref{fig:ave}, as ``E''=engine). At 
smaller $\tau$, the analysis of the heats reveals the existence of 
three regimes, marked on the Figure as ``D'', ``R'' and again ``D''. 
In the ``R'' regime the system acts as a refrigerator, i.e. consumes 
work to push heat from $T_L$ to $T_H$. In the ``D'' phases, the heat 
flow is the standard one (from $T_H$ to $T_L$), even if work is 
consumed: however, the rate of heat transfer $\langle Q_2 
\rangle/\tau$ is higher than in the ``E'' phase, and therefore the 
machine acts as a more efficient heat sink, similar to dissipating 
fans. At a time $\tau_{res}<\tau^*$ we notice the presence of a 
maximum in $\langle W \rangle$: it is of the order of magnitude of 
the adiabatic limit, but with opposite sign. At smaller $\tau \to 0$ 
the consumed work goes to $0$. Let us note that the relevant 
timescales emerged from this analysis are in fair agreement with the 
characteristic relaxation times computed in a simple Langevin model of 
this system, see below.

\subsection{Power and efficiency} 
\begin{figure}[!hbtp]
\centering
\includegraphics[width=\columnwidth]{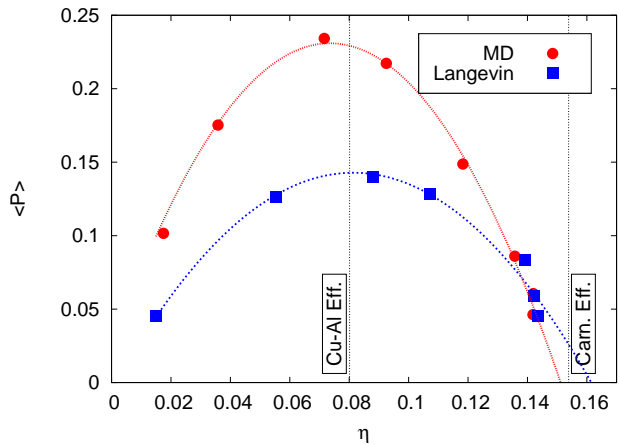}
\caption{Average power versus efficiency for MD (red
  curve) and for the reduced ``3V'' model, Eq.~\eqref{eq:3v}
  (blue curve). The Curzon-Ahlborn estimate and the Carnot efficiency
  are also indicated. }\label{fig:eff}
\end{figure}

Measures of the average developed power $\langle P \rangle=-\langle W
\rangle/\tau$ are reported in Fig.~\ref{fig:eff} (red curve) in the
engine phase $\tau>\tau^*$.  Those measures are given as a function of
the efficiency $\eta=-\langle W \rangle/\langle Q_2 \rangle$, which is
monotonically increasing with $\tau$. At a time around $\sim
500$, a maximum is observed in $\langle P \rangle$, whose associated
efficiency is found to be slightly smaller than the Curzon-Ahlborn
(CA) estimate \cite{curzon1975efficiency} $\eta_{CA} = 1- \sqrt{T_L/T_H}$. We recall that
the CA estimate is based upon an endo-reversible model of Carnot
engine where the only entropy changes (even at finite times) occur in
the heat transfers. Recently a wider hypothesis for the CA formula
has been investigated, which seems to be
``symmetric dissipation'', i.e. equal entropy production rates during
the two isothermals~\cite{Esposito2010}. It is likely that our choice of
values for $T_H$ and $T_L=0.85\, T_H$ puts us close to that
scenario. Nevertheless it is remarkable to recover a result usually
derived through macroscopic theories, i.e. without fluctuations, in a
small system such as our molecular model.

\subsection{Fluctuations} 
\begin{figure}[htbp]
\centering
\includegraphics[width=\columnwidth]{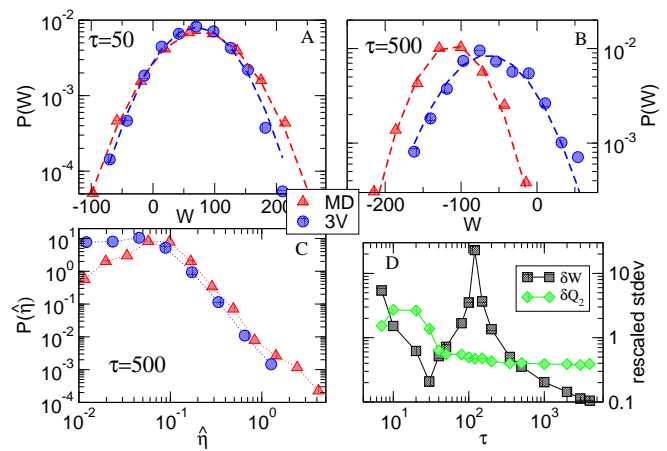}
\caption{Study of fluctuations. A) and B): Pdf of the work in a cycle
  for two different values of $\tau$ (``E'' and ``D'' regimes), from
  the MD and from the reduced ``3V'' model Eq.~\eqref{eq:3v}.  C) Pdf
  of the fluctuating efficiency in a cycle at $\tau=500$. D) rescaled
  stdev (see text) of $P(W)$ and of $P(Q_2)$ as a function of
  $\tau$. The statistics in this Figure is obtained from $2000$
  cycles.}\label{fig:fluct}
\end{figure}

In small systems, fluctuations are hardly negligible~\cite{ld2014}. In
Fig.~\ref{fig:fluct} (A and B, red curves), we display the behavior of
fluctuations of work $W$ integrated in a cycle for two different
regimes, at $\tau=50<\tau^*$ and $\tau=500>\tau^*$. Deviation from a
Gaussian behavior are small, indicating that $N$, even if finite, is
large enough to expect the validity of the central limit theorem.  Interestingly the measure
of the standard deviation (stdev) $\sigma_W$ rescaled by the average value $\delta W=\sigma_W/|\langle
W \rangle |$ (black curve in Fig.~\ref{fig:fluct}D) shows that $\delta
W \ll 1$ close to $\tau_{res}$ and $\delta W \gg 1$ at the stall time
$\tau^*$. The relative stdev for the heat, $\delta Q_2=\sigma_{Q_2}/|\langle Q_2 \rangle|$ behaves much more
regularly. It is also interesting to analyze the fluctuation of the
``fluctuating efficiency'', i.e. $\hat{\eta}=-W/Q_2$ measured in a
single cycle, see Fig.~\ref{fig:fluct}C (restricted to positive
values), which displays a long tail for values
larger than the average~\cite{Verley2014}.

\section{Coarse-grained description} 
\label{sec: 3}
In order to make contact with stochastic
thermodynamics~\cite{Seifert2005}, which is a useful framework for
small systems, we need a coarse-grained description with few
relevant (slowly-changing) variables. The contribution of the fast 
degrees of freedom is in the noise. Reasonable candidates are: the 
position of the piston $X$, its velocity $V$ and the estimate of the 
instantaneous ``temperature'' of the gas $T(t)=m/N \sum_{i=1}^N 
v_i^2(t)$. The time evolution of these observables can be determined 
by computing the average rate of collision occurring between the 
particles of the gas and the walls of the container. Here, at any $t$, 
we assume the gas to be homogeneously distributed in the interval 
$(0,X(t))$ and each particle to have a velocity $v$, given by a 
Maxwell-Boltzmann distribution $\rho_T(v)$ at the temperature $T(t)$. 
In addition we use the fact that the collisions between the gas 
particles and the piston are elastic and that a particle that collides 
with the thermal wall gets a new velocity $v'$ distributed according 
to a Maxwellian distribution $\rho_{o} (v')$. Taking into account the 
contributions of the external force and the collision, we have that 
the average derivative of the velocity of the piston $\langle 
\dot{V}\rangle =\lim_{\Delta t \to 0} \langle\Delta V\rangle/\Delta t$  
is
\begin{equation}\label{eq: piston}
\langle \dot{V}\rangle=
-\frac{F}{M}+\frac{N}{X}\int\,dv \,\,\frac{2m(v-V)^2}{(M+m)}\Theta(v-
V) \rho_T(v).
\end{equation}

On the other hand, $\langle \dot{T}(t)\rangle$ is the sum of two terms 
coming from the collisions with the piston
\begin{equation}\label{eq: temperature1}
\left .\dot{\langle T \rangle}\right|_{coll}=\frac{m}{X}\int\,dv \,\,(v'^2-v^2) |v-V|\Theta(v-V)\rho_T(v),
\end{equation}
where $v'$ is the velocity after an elastic collision, and the
interaction of the gas with the thermostat
\begin{equation}\label{eq: temperature2}
\left.\dot{ \langle T \rangle}\right|_{ther}=\frac{m}{X}\int\,dv\,dv' (v'^2-v^2)\,|v|\,\Theta(-v)\,\rho_{T}(v)\rho_{o}(v').
\end{equation}

\begin{figure}[!tbp]
\centering \includegraphics[width=\columnwidth]{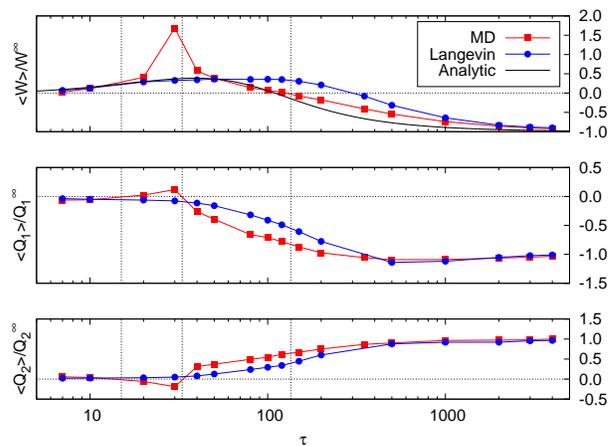}
\caption{Average work and heats per cycle. Comparison between the
  MD, the $3$-variables model, Eq.~\eqref{eq:3v},
  and the analytical solution in Eq.~\eqref{anwork} rescaled by their asymptotic values.}\label{fig:theory}
\end{figure}

In order to reduce the dynamics to a linear Langevin equation we 
assume the fluctuations of $X,V$ and $T$ to be small (such assumptions
are reasonable if $N\gg1$ and $M\gg m$) and expand Eqs. (\ref{eq: 
piston}), (\ref{eq: temperature1}) and (\ref{eq: temperature2}) up to 
the first order around the equilibrium values $X_{eq}=N T_{o}/F$, 
$V_{eq}=0$ and $T_{eq}=T_o$. The linearity of the equation is, on one 
hand,  inspired by the gaussianity of pdfs, and, on the other, it is a 
useful assumption that allows simple computations. The stochastic part 
is obtained by adding the gaussian noise terms with amplitudes 
determined by imposing that the variances of the variables coincide 
with those computed within the canonical 
ensemble~\cite{cerino2014fluctuations}. This yields
\begin{eqnarray}\label{eq:3v}
\dot{X}&=&V, \nonumber\\
\dot{V}&=&-k (X-X_{eq}) -\gamma V +\mu (T-T_{eq}) \nonumber + \sqrt{\frac{2\gamma T_o}{M}} \xi_1,\\
\dot{T}&=&  - \frac{2 M T_o}{N} \mu V-\alpha (T- T_{eq})+\sqrt{\frac{4\alpha T_o^2}{N}}\xi_2,
\end{eqnarray}
where $\xi_1$ and $\xi_2$ are independent white noises $\langle \xi_i
\rangle =0$, $\langle
\xi_i(t)\xi_j(t')\rangle=\delta_{ij}\delta(t-t')$, ${k(t)=F(t)^2/[M
    NT_o(t)]}$, ${\gamma(t)=2F(t)\sqrt{2m/[ M^2\pi T_o(t)]}}$,
${\mu(t)=F(t)/[M T_o(t)]}$ and ${\alpha(t)=F(t)\sqrt{2/ [mN^2\pi
T_o(t)]}}$.  A numerical study of the ``3 variables'' (3V) model
in Eq.~\eqref{eq:3v} reveals a fair agreement with our main
observations. In Fig.~\ref{fig:theory} the average values per cycle of
work and heats are compared with those obtained in the original MD: we
define $\dot{W}$ and $\dot{Q}$ as in MD, with $E(t)=NT(t)/2+MV(t)^2/2
+F(t)X(t)$. The maximum and the inversion of the average work are
fully reproduced, but with significant shifts of the values of $\tau$
where they occur. Indeed, a more detailed analysis (not reported here) 
has identified the relevance of two additional variables: taking into
account the position and the velocity of the center of mass of the
gas, it is possible to achieve a better agreement with the
MD. Unfortunately, the parameters for such a ``5 variables'' model can
only be obtained by fitting the MD data.  Notwithstanding its degree
of approximation, the $3V$ model gives a fair account of efficiency at
maximum power, see Fig.~\ref{fig:eff}, as well as of fluctuations, see
Fig.~\ref{fig:fluct}A and B, which have a very similar Gaussian shape
and width. The overall shape of the efficiency fluctuations' pdf
(Fig.~\ref{fig:fluct}C) is also reproduced. The eigenvalues of the
dynamic's matrix in Equations~\eqref{eq:3v} give also access to 
typical timescales. For instance, for $F=200$, $T=12$, $M=100$, $m=1$ 
and $N=500$, the eigenvalues read $\lambda_1 \approx -0.02$ and
$\lambda_{2,3}\approx-0.50\pm i\,\,0.10$ leading to three
characteristic timescales to compare with the total duration of the
cylce: $\tau_1=4/|\lambda_1|\approx 170$, $\tau_2=4/|
\Re(\lambda_{2,3})|\approx 8 $ and $\tau_3=4/|
\Im(\lambda_{2,3})|\approx40$. Remarkably, the order of magnitude of 
the relevant timescales in the MD system is correctly reproduced by 
the eigenvalues of the equilibrium dynamic's matrix (see Fig. 
\ref{fig:ave} and Fig. \ref{fig:theory}). A detailed study of the 3V 
model is out of our present scope, but certainly deserves future 
investigation.
\section{A solvable toy model}
\label{sec: 4}
In spite of its apparent simplicity, it is not easy to derive 
analytical results for the 3V model in a cycle of the external 
parameters. Here we show that the qualitative dependence of $\langle W
\rangle$ on the total time of the cycle $\tau$ can be obtained in
a simplified version of Eq. (\ref{eq:3v}), where we set the 
temperature $T(t)$ to be equal to the temperature of the
thermostat $T_o(t)$ at every time $t$. In addition, we assume
the parameters to vary in the following form ($\omega=2\pi/\tau$) 
\begin{eqnarray}\label{eq: parametri} 
f(t)&=&f_0(1+\epsilon\cos(\omega t)),\nonumber\\
T(t)&=&T_0(1+q\epsilon\sin(\omega t)),
\end{eqnarray}
where ${f(t)=F(t)/M}$, $\epsilon \ll 1$ and ${q\sim\mathcal{O}(1)}$:
we set ${f_0=2}$, $T_0=12$, $\epsilon=0.1$ and $q=0.8$. In the
adiabatic limit, this simplified cycle (an approximation of the 
Ericsson protocol, see Fig. \ref{fig: parameters}) produces a work not 
very different from the one of the Ericsson cycle.  Passing to average
values ($Y=\langle X \rangle$) we obtain the equation (see Appendix 
\ref{app: B}):
\begin{equation}\label{eq: init}
\ddot{Y}+k(t) Y +\gamma(t) \dot{Y} = f(t).
\end{equation}

The homogeneous solution associated to Eq. (\ref{eq: init}) goes to 
zero in the long time limit: therefore, since we are interested in the
asymptotic stationary solution, we will focus only on the non-homogeneous solution. This will be done by expanding all the terms in Eq. (\ref{eq: init}) in powers of $\epsilon$. In particular, since
$Y(t)=Y_0(t)+\epsilon Y_1(t) +\mathcal{O}(\epsilon^2)$,
by solving Eq.~\eqref{eq: init} for $\epsilon \to 0$ one gets
$Y_0(t)=\frac{N T_0}{F_0}$, and
\begin{equation}
\ddot{Y}_1+\omega_0^2 Y_1 +\nu \dot{Y}_1=-f_0\cos \omega t +f_0 q \sin \omega t,
\end{equation}
where $\omega_0^2= F_0^2/(MN T_0)$ and $\nu=2f_0\sqrt{2m/(\pi T_0)}$.
The asymptotic solution is 
\begin{equation}
Y_1(t)=A(\omega)[ \cos(\omega t +\phi(\omega))-q \sin (\omega t +\phi(\omega))],
\end{equation} 
\begin{eqnarray}
A(\omega)&=&\frac{-f_0}{\sqrt{\left(\omega_0^2-\omega ^2\right)^2+\nu ^2\omega ^2}},\nonumber \\
\phi (\omega)&=&\arctan\left(\frac{\nu  \omega }{\omega_0^2-\omega^2}\right).
\end{eqnarray}
The work performed over a cycle of the parameters of total time
$\tau$ can be now expressed in a simple way
\begin{eqnarray} \label{anwork}
W(\tau)&=&-Mf_0\frac{2\pi}{\tau} \epsilon\int_0^\tau  dt\,\, (Y_0+\epsilon Y_1(t))  \sin \left( 2\pi \frac{t}{\tau}\right)=\nonumber\\
=- Mf_0 \pi &\epsilon^2&A\left(\frac{2\pi}{\tau}\right)\left[\sin\phi\left(\frac{2\pi}{\tau}\right) -q\cos\phi\left(\frac{2\pi}{\tau}\right)\right].
\end{eqnarray}
In Fig.~\ref{fig:theory} (black curve) it is seen that this result,
when normalized to its adiabatic value, compares quite well, in spite 
of the many approximations introduced to obtain Eq.~\eqref{anwork}, 
with the average work performed by the MD system and the 3V model, 
recovering the change of sign at value not far from $\tau^*$ and a 
maximum at smaller values. Computing the heat transfers is a more 
difficult task, as it requires a solution of Eq.~\eqref{eq: init} at 
order $\mathcal{O}(\epsilon^2)$.

\section{Concluding remarks} 
Summarizing, we have introduced a new model for a heat engine where
fluctuations (due to small $N$) and finite power (due to small $\tau$)
are observed. In the past a great attention has been given to
extremely simplified models, typically in the form of single molecules
or colloids, or Markov chains inspired to biochemical reactions. Here
we propose to move towards a higher level of complexity, and possibly
realism, suggesting a new test-ground for statistical mechanics of
small systems out-of-equilibrium. Our system reveals non-trivial
features, such as several working regimes (engine, refrigerator, heat
sink) tuned by simply controlling $\tau$. Notwithstanding its rich
phenomenology, the model admits a coarse-grained description in terms 
of three linearly coupled Langevin equations. Further investigation of 
this reduced model are in order, in particular of heat, work and 
efficiency fluctuations~\cite{Verley2014}.


\begin{acknowledgments}
We acknowledge useful discussions with T. Sano and D. Villamaina. Our 
work is supported by the ``Granular-Chaos'' project, funded by the 
Italian MIUR under the FIRB-IDEAS grant number RBID08Z9JE.
\end{acknowledgments}

\appendix
\section{Details on the derivation of the Langevin Equation}
\label{app: A}
In order to get a linear Langevin equation from kinetic theory we
start from the conditional equilibrium distribution with fixed values
of the macroscopic variables $X,V,T$, and then determine the average
number of particles that, in the unit time $\Delta t$, collide with 
the piston or with the thermal wall. Using Eq. (\ref{eq: elastic}) and
(\ref{eq: thermostat}), one can determine post-collisional velocities
and, accordingly, the average change of $V$ and $T$, over a time
$\Delta t$. In the following, to simplify the notation, we denote with
$\langle \cdot \rangle$ the conditional average $\langle \cdot
|X,V,T\rangle$.

The equation for $X$ reads $\dot{X}=V$. The total average
force acting on the piston due to collisions is:
\begin{eqnarray}\label{eq: vdot}
&&\lim_{\Delta \to 0} \frac{\langle \Delta V\rangle_{coll}}{\Delta t}=\nonumber\\
&=&\frac{N}{X}\int_0^\infty dv \,\frac{2m}{m+M}(v-V)|V-v|\rho_T(v)\Theta(V-v)=\nonumber\\
&=&\frac{2Nm}{(M+m)X}\sqrt{\frac{m}{2\pi T}}\int_V^{\infty}dv\,\,(v-V)^2e^{-m\frac{v^2}{2 T}}=\nonumber\\
&=&\frac{N}{(m+M)X} \Bigg(\left(T+m V^2\right) \text{erfc}\left(\sqrt{\frac{m}{2T}} V\right)+\nonumber\\
&&-~~\sqrt{\frac{2m TV^2}{\pi}} e^{-\frac{m V^2}{2 T}}\Bigg),
\end{eqnarray}
where $\text{erfc}(x)=2/\sqrt{\pi}\int_x^\infty dt \exp(-t^2)$. To obtain the total force, a term $-F/M$ must be added. The  elastic 
collisions with the piston also affect the kinetic energy of the gas, 
through the term

\begin{widetext}
\begin{eqnarray}\label{eq: tdot1}
\left .\dot{\langle T \rangle}\right|_{coll}&=&\frac{m}{X}\int\,dv \,\,(v'^2-v^2) |v-V|\Theta(v-V)\rho_T(v)=\nonumber\\
&=&\frac{m}{X}\sqrt{\frac{m}{2 \pi T}}\int_V^{\infty} dv\,\left[\left(v+2\frac{M}{m+M}(V-v)\right)^2-v^2\right](v-V)e^{-m \frac{v^2}{2T}}=\nonumber\\
&=&-\frac{2 M}{(m+M)^2 X}\left(\sqrt{\frac{2m T}{\pi }} \left(2 T-M V^2\right)e^{-\frac{m V^2}{2 T}}+ V \left(M T+m M V^2-2 m T\right) \text{erfc}\left(\sqrt{\frac{m}{2T}}V\right)\right).
\end{eqnarray}
Finally, the average change of temperature in a time interval $\Delta 
t$, due to the collision with the thermal wall is simply given by the 
term 

\begin{eqnarray}\label{eq: tdot2}
\left.\dot{ \langle T \rangle}\right|_{ther}&=&\frac{m}{X}\int\,dv\,dv' (v'^2-v^2)\,|v|\,\Theta(-v)\,\rho_{T}(v)\rho_{o}(v')=\nonumber\\
&=&-\frac{m^2}{XT_o} \sqrt{\frac{m}{2\pi  T}}\int_{-\infty}^0 dv\int_0^\infty dv'\,\,\left(v'^2-v^2\right)v v' e^{-m \frac{v^2}{2T}}e^{-m \frac{v'^2}{2T_o}}\nonumber=\\
&=&\sqrt{\frac{2}{\pi m}}\frac{\sqrt{T}(T_o-T)}{X}.
\end{eqnarray}

\end{widetext}

The equilibrium value of $X,V$ and $T$ for which $\langle\dot{X}\rangle=0$, $\langle\dot{V}\rangle=0$ and $\dot{T}=\left .\dot{\langle T \rangle}\right|_{coll}+\left.\dot{ \langle T \rangle}\right|_{ther}=0$ are 
\begin{eqnarray}
X_{eq}&=&N\frac{T_o}{F},\\
V_{eq}&=&0,\\
T_{eq}&=&T_o,
\end{eqnarray}
where terms $\mathcal{O}(m/M)$ are neglected. We can obtain a linear equation by expanding the expressions above up to first order around the equilibrium values: this can be done only if fluctuations are small, i.e. when $N\gg 1$ and $M\gg m$.
The sum of Eq. (\ref{eq: vdot}) and $-F/M$ yelds
\begin{equation}
\langle \dot{V} \rangle =-k (X-X_{eq}) -\gamma V +\mu (T-T_{eq}),
\end{equation}
with $k=F^2/MNT_o$, $\gamma=2F\sqrt{2m/(M^2\pi T_o)}$ and  $\mu=F/(MT_o)$ . Similarly the sum of Eq. (\ref{eq: tdot1}) and Eq. (\ref{eq: tdot2}) yelds
\begin{equation}
\langle \dot{T}\rangle =  - \frac{2 M T_o}{N} \mu V-\alpha (T- T_{eq}),
\end{equation}
with $\alpha=F\sqrt{2/(m N^2\pi T_o)}$. The coefficients $k,\gamma,\mu$ and $\alpha$ vary in time according to the time evolution of $F$ and $T_o$. In order to take into account the fluctuations of this variables one must add three independent gaussian terms $\xi_X\equiv \xi_0$, $\xi_V\equiv\xi_1$ and $\xi_T\equiv\xi_2$, with an appropriate weight matrix $b_{ij}$ with $i,j=0,1,2$. In this particular case the matrix is diagonal, $b_{ii}=b_i$, with $b_0=0$, $b_1=\sqrt{2\gamma \sigma^2_V}$ and $b_2=\sqrt{2\gamma \sigma^2_T}$. The final form of the linear Langevin equation thus reads
\begin{eqnarray}\label{eq:3vapp}
\dot{X}&=&V, \nonumber\\
\dot{V}&=&-k (X-X_{eq}) -\gamma V +\mu (T-T_{eq}) \nonumber + \sqrt{\frac{2\gamma T_o}{M}} \xi_1,\\
\dot{T}&=&  - \frac{2 M T_o} {N} \mu V-\alpha (T- T_{eq})+\sqrt{\frac{4\alpha T_o^2}{N}}\xi_2.
\end{eqnarray}
Let us note that this equation, with fixed $F$ and $T_o$, satisfies detailed balance \cite{gardiner1985handbook}. 

\section{ Details on the analytic solution of the toy model}
\label{app: B}
Let us consider Eq. (\ref{eq: init}):
\begin{equation}\label{eq: initapp}
\ddot{Y}+k(t) Y +\gamma(t) \dot{Y} = f(t),
\end{equation}
where
\begin{eqnarray}
\gamma(t) &=&\frac{2F(t)}{M}\sqrt{\frac{2m}{\pi T_o(t)}}, \nonumber\\
k(t) &=& \frac{F(t)^2}{MNT_o(t)},
\end{eqnarray}
and $f(t)=F(t)/M$. In the Ericsson cycle $F$ and $T$ depend on time in a too complicated manner in order to perform analytic calculations. Therefore, in order to obtain an explicit result, in the following we will assume
\begin{eqnarray}
f(t)&=&f_0(1+\epsilon\cos(\omega t)),\nonumber\\
T_o(t)&=&T_0(1+q\epsilon\sin(\omega t)),
\end{eqnarray}
where $\omega=2\pi/\tau$, $\epsilon\ll1$ and $q\sim\mathcal{O}(1)$. We will now sketch the derivation of the non-homogeneous solution of Eq. (\ref{eq: initapp}) as an asymptotic expansion in $\epsilon\ll 1$. Let us expand in power of $\epsilon$ all the coefficients appearing in Eq. (\ref{eq: initapp}) up to $\mathcal{O}(\epsilon)$
\begin{eqnarray}
k(t)&\approx&\frac{M f_0^2}{NT_0}\left[1+(2\cos \omega t-q\sin\omega t )\epsilon\right]\\
\gamma (t)&\approx&\sqrt{\frac{4f_o^2m}{\pi T_0}}\left[1+\left(\cos \omega t- \frac{q}{2}\sin\ \omega t\right)\epsilon\right] \nonumber\\
Y(t)&\approx&Y_0(t)+\epsilon Y_1(t).
\end{eqnarray} 

Plugging these expressions into Eq. (\ref{eq: initapp}), for $\epsilon=0$ one gets

\begin{equation}
\ddot{Y}_0+\frac{M f_0^2}{N T_0} Y_0 +2f_0\sqrt{\frac{2m}{\pi T_0}} \dot{Y}_0 = f_0,
\end{equation}
leading to 
\begin{equation}
Y_0(t)=\frac{N T_0}{F_0}.
\end{equation}
At the following order $\mathcal{O}(\epsilon)$, Eq. (\ref{eq: initapp}) gives
\begin{eqnarray}
\ddot{Y}_1+\frac{M f_0^2}{N T_0}(Y_1 +(2\cos \omega t-q\sin\omega t )Y_0)+\nonumber\\
 2f_0\sqrt{\frac{2m}{\pi T_0}}\left(\dot{Y}_1+\left(\cos \omega t - \frac{q}{2}\sin\ \omega t\right)\dot{Y}_0\right)=\nonumber\\= f_0 \cos \omega t
\end{eqnarray}
or, if we plug the value of $Y_0$,
\begin{equation}
\ddot{Y}_1+\omega_0^2 Y_1 +\nu \dot{Y}_1=-f_0\cos \omega t +f_0 q \sin \omega t,
\end{equation}
where $\omega_0=\sqrt{M f_0^2/(NT_0)}$ and $\nu=2f_0\sqrt{2m/(\pi T_0)}$. A solution of this equation can be found in the form
\begin{equation}
Y_1(t)=A(\omega) (\cos(\omega t +\phi(\omega))+ q\sin (\omega t +\phi(\omega))),
\end{equation}
where
\begin{eqnarray}
A(\omega)&=&\frac{-f_0}{\sqrt{\left(\omega_0^2-\omega ^2\right)^2+\nu ^2\omega ^2}},\\
\phi (\omega)&=&\arctan\left(\frac{\nu  \omega }{\omega_0^2-\omega^2}\right).
\end{eqnarray}
\bibliography{mergedbiblio}

\end{document}